\documentclass{PoS}

\title{The nature of 19 X--ray sources detected with {\it INTEGRAL}}

\ShortTitle{Identification of 19 IGR sources}

\author{N. Masetti$^1$, P. Parisi$^{1,2}$, E. Jim\'enez-Bail\'on$^3$, 
L. Bassani$^1$, A. Bazzano$^4$, A.J. Bird$^5$, V. Chavushyan$^6$, 
A.J. Dean$^5$, G. Galaz$^7$, R. Landi$^1$, A. Malizia$^1$, D. 
Minniti$^{7,8}$, L. Morelli$^9$, E. Palazzi$^1$, 
F. Schiavone$^1$, J.B. Stephen$^1$ and P. Ubertini$^4$\\
$^1$INAF/IASF di Bologna, Italy\\
$^2$Universit\`a di Bologna, Italy\\
$^3$Universidad Aut\'onoma de M\'exico, Mexico DF\\ 
$^4$INAF/IASF di Roma, Italy\\ 
$^5$University of Southampton, UK\\ 
$^6$INAOE, Puebla, Mexico\\ 
$^7$Pontificia Universidad Cat\'olica, Santiago, Chile\\
$^8$Specola Vaticana, Citt\`a del Vaticano\\
$^9$Universit\`a di Padova, Italy\\
        E-mail: \email{masetti@iasfbo.inaf.it}}

\abstract{Since its launch on October 2002, the {\it INTEGRAL} satellite 
has revolutionized our knowledge of the hard X--ray sky thanks to its 
unprecedented imaging capabilities and source detection positional 
accuracy above 20 keV. Nevertheless, many of the newly-detected sources in 
the {\it INTEGRAL} sky surveys are of unknown nature. The combined use of 
available information at longer wavelengths (mainly soft X--rays and 
radio) and of optical spectroscopy on the putative counterparts of these 
new hard X--ray objects allows pinpointing their exact nature. Continuing 
our long-standing program running since 2004, here we report the 
classification, through optical spectroscopy, of 19 more unidentified 
or poorly studied high-energy sources detected with the IBIS instrument 
onboard {\it INTEGRAL}.}

\FullConference{The Extreme and Variable High Energy Sky - extremesky2011,\\
		September 19-23, 2011\\
		Chia Laguna (Cagliari), Italy}

\begin{document}

\section{Introduction}

One of the main objectives of the {\it INTEGRAL} \cite{w03} is a regular 
survey of the whole sky in the hard X--ray band. This makes use of the 
unique imaging capability of the IBIS instrument \cite{u03} which allows 
the detection of sources at the mCrab level with a typical localization 
accuracy of a few arcmin. The latest (4$^{\rm th}$ one) IBIS catalogue 
published up to now by \cite{b10} shows that, among more than 700 sources 
detected above 20 keV, a substantial fraction ($\sim$30\%) had no obvious 
counterpart at other wavelengths and therefore could not yet be associated 
with any known class of high-energy emitting objects.

With the task of identifying them, our group started in 2004 an optical
spectroscopy program which used data from ten telescopes worldwide up
to now and which proved extremely successful, leading to the
identification of about 150 {\it INTEGRAL} objects\footnote{The list 
of {\it INTEGRAL} sources identified through optical and near-infrared 
spectroscopy can be found at {\tt 
http://www.iasfbo.inaf.it/extras/IGR/main.html}}, that is about three
quarters of the unidentified sources sample of the first three IBIS
catalogues.

It is found that more than half of the sources identified by us are 
Active Galactic Nuclei (AGNs), followed by X--ray binaries ($\sim$30\%) 
and Cataclysmic Variables (CVs; $\sim$13\%); see \cite{m10} and 
references therein.

Here we present the continuation of this follow-up and identification work 
on {\it INTEGRAL} objects with the use of data collected in more than one 
year, between March 2010 and May 2011.

\section{Sample selection}

Using the criterion applied in our past works (see \cite{m10} and 
references therein), we positionally cross-correlated the unidentified 
sources belonging to recently published IBIS surveys 
(\cite{bi07,b10,k10,p11}) with available X--ray ({\it ROSAT}, {\it 
XMM-Newton}) and radio (NVSS, SUMSS, MGPS) on-line catalogues, and with 
archival X--ray observations ({\it Swift}, {\it Chandra}) to reduce the 
arcminute-sized source error circle to a radius of better than a few 
arcsec, thus pinpointing the putative optical counterpart.

This process eventually allowed us to select 19 new candidate counterparts 
to be observed through optical spectroscopy. The list of selected objects 
is reported in the first column of Table 1.

\section{Caveats}

While it is known \cite{s06} that the presence of a single, bright soft 
X--ray object within the IBIS error circle indicates that it is with a 
high probability the lower-energy counterpart of the corresponding {\it 
INTEGRAL} source, the same cannot be said with high confidence for radio 
objects (see however \cite{m11}). Thus, for the IBIS sources which were 
singled out using catalogues at wavelengths longer than soft X--rays, we 
caution the reader that the association with the selected optical object 
should await confirmation via a pointed observation with a soft X--ray 
satellite such as {\it XMM-Newton}, {\it Chandra} or {\it Swift}.

We also stress that in a few cases the putative soft X--ray (and optical) 
counterpart lies outside the nominal IBIS 90\% confidence level error 
circle (but within the 99\% one). 

To remark the above uncertainties, we put an asterisk beside the source 
name in Table 1.

\section{Observations}

In this work, 6 telescopes located in Northern and Southern Hemisphere 
observatories were used; moreover, a spectrum from the 6dF on-line 
archive \cite{j04} was also employed. In the following, the list of the 
telescopes used for the classifications reported in Table 1 is given:

\begin{itemize}

\item 1.5m telescope at CTIO, Chile;

\item 1.5m ``Cassini" telescope in Loiano, Italy;

\item 1.8m ``Copernico" telescope in Asiago, Italy;

\item 2.1m telescope in San Pedro M\'artir, Mexico;

\item 3.58m Telescopio Nazionale Galileo, Canary Islands, Spain;

\item 3.9m AAT telescope of the Anglo-Australian Observatory, Siding 
Spring, Australia (for the 6dF spectrum);

\item 10.4m Gran Telescopio Canarias, Canary Islands, Spain.

\end{itemize}

\section{Results}

Table 1 reports, for each selected source, the corresponding 
classification obtained through optical spectroscopy, its redshift and 
the name of the telescope with which the observation was performed; in 
Fig. 1 the spectra of the optical counterparts of some of these 19 
selected IBIS sources are shown.

\begin{figure}
\vspace{-2cm}
\hspace{-1cm}
\includegraphics[width=18cm]{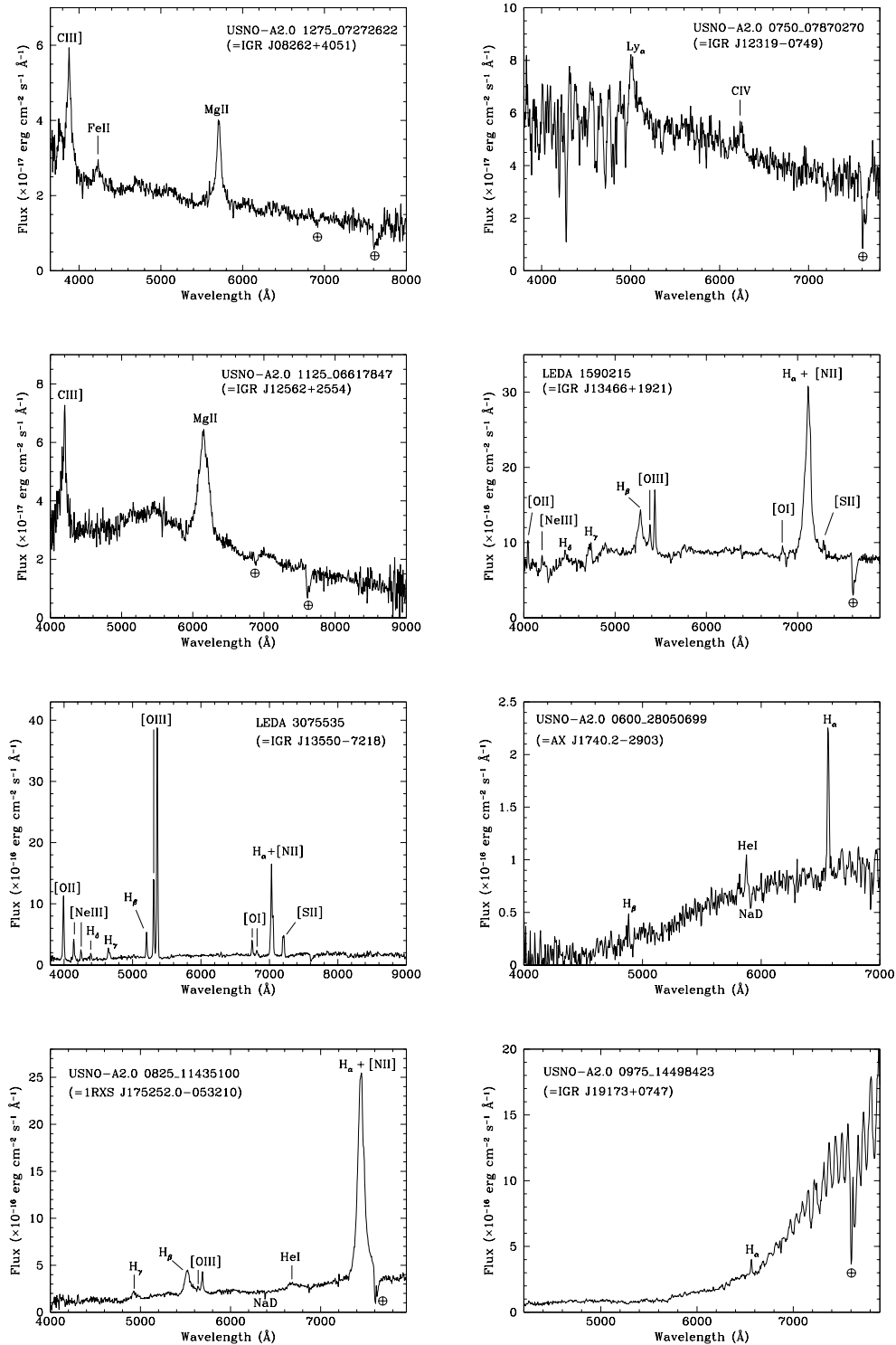}
\vspace{-4.5cm}
\caption{Optical spectra of the counterparts of 8 objects belonging to 
the sample of 19 {\it INTEGRAL} sources identified in this work and 
listed in Table 1. These data allowed us to securely determine the nature 
and the redshift of these objects through inspection of their absorption 
and emission spectral features. The spectra are not corrected for the 
intervening Galactic absorption. For each spectrum the main spectral 
features are labeled. The symbol $\oplus$ indicates atmospheric telluric 
absorption bands.}
\end{figure}

\begin{table}%[ph]
\caption{The sample of 19 {\it INTEGRAL} objects identified in this work 
together with their classification, redshift and telescope with which the 
identification was obtained. The asterisks indicate sources for which an 
X--ray position either is still not available or lies outside the 90\% 
IBIS error circle (see Sect.~3).}
\begin{center}
\begin{tabular}{llll}
\noalign{\smallskip}
\hline
\hline
\noalign{\smallskip}
Object name & Class & redshift & Telescope \\ 
\noalign{\smallskip}
\hline
\noalign{\smallskip}
IGR J03184$-$0014$^*$   & Sy1.9      & 0.330 & GTC    \\
IGR J03249+4041         & double Sy2 & 0.049 & SPM    \\
RX J0525.3+2413         & CV         & 0     & SPM    \\
IGR J05255$-$0711$^*$   & QSO        & 0.634 & TNG    \\
IGR J06073$-$0024$^*$   & QSO        & 1.028 & TNG    \\
IGR J06523+5334         & Sy1.2      & 0.301 & SPM    \\
IGR J08262+4051         & QSO        & 1.038 & TNG    \\
IGR J12319$-$0749       & Blazar     & 3.12  & TNG    \\
IGR J12562+2554$^*$     & QSO        & 1.199 & GTC    \\
IGR J13466+1921         & Sy1.2      & 0.085 & SPM    \\
IGR J13550$-$7218       & Sy2        & 0.071 & CTIO   \\
IGR J14385+8553$^*$     & Sy1.5      & 0.081 & Asiago \\
IGR J17197$-$3010$^*$   & Symbiotic  & 0     & SPM    \\
AX J1740.2$-$2903       & CV         & 0     & SPM    \\
1RXS J175252.0$-$053210 & Sy1.2      & 0.136 & SPM    \\
IGR J18371+2634$^*$     & RS CVn?    & 0     & Loiano \\
IGR J19173+0747         & HMXB       & 0     & SPM    \\
IGR J19491$-$1035       & Sy1.2      & 0.024 & 6dF    \\
IGR J20450+7530         & Sy1        & 0.095 & Loiano \\
\noalign{\smallskip}
\hline
\noalign{\smallskip}
\end{tabular}
\end{center}
\end{table}

As one can see from Table 1, most objects (14 out of 19) are AGNs, with 
a large majority (12) of Type 1 (i.e. broad-line) ones. Among them, 5 
lie at large ($z >$ 0.5) redshift: this confirms the trend that new, 
deeper IBIS surveys are able to detect more distant AGNs. In particular, 
we were able to identify IGR J12319$-$0749 (at $z$ = 3.12) as the second 
most distant object ever detected by {\it INTEGRAL} up to now after the
blazar IGR J22517+2218, which lies at redshift $z$ = 3.668 (\cite{ba07} 
and references therein).

Within our AGN identifications we moreover refine the classifications of 
IGR J06523+5334 and IGR J19491$-$1035 made by \cite{h11} and \cite{k11}, 
respectively. We also confirm the double Seyfert 2 nature of source IGR 
J03249+4041 first suggested by \cite{l10}.

Only 5 of our identifications are instead of Galactic nature: two of them 
are (likely magnetic) CVs, one is a Symbiotic Star, one is a magnetic 
flare star (possibly of RS CVn type), and one is a High-Mass X-ray Binary 
(HMXB). We confirm the CV nature of sources RX J0525.3+2413 and AX 
J1740.2$-$2903 as first proposed by \cite{t07} and \cite{hg10}, 
respectively; we also note that the spectra of these two objects (along 
with that of the HMXB IGR J19173+0747) appear substantially reddened. We 
further stress that the optical spectrum of AX J1740.2$-$2903 (left panel 
of third row from top in Fig. 1) allows us to exclude the possibility that 
it is a Symbiotic X--ray binary as proposed by \cite{f10}, as in this case 
we would have seen the optical spectral continuum typical of a late-type 
giant star (see e.g. \cite{m06}).

\section{Conclusions}

Here we presented the precise identification of a set of 19 sources of 
unknown or uncertain nature and which were detected by IBIS onboard {\it 
INTEGRAL}. We found that nearly three quarters of them have an 
extragalactic nature, with redshifts in the range 0.024--3.12, while the 
remaining ones are Galactic sources. It is noteworthy that the majority of 
the latter ones are (possibly magnetic) CVs or Symbiotic stars.

These preliminary results further confirm the {\it INTEGRAL} capabilities 
of detecting AGNs and hard X-ray emitting CVs.

\section*{Acknowledgments}

This research has made use of the SIMBAD database operated at CDS, 
Strasbourg, France, of the NASA/GSFC's HEASARC archive, and of the 
HyperLeda catalogue operated at the Observatoire de Lyon, France.
%We thank the referee for useful indications.
NM acknowledges financial contribution from the ASI-INAF agreement No. 
I/009/10/0. PP has been supported by the ASI-INTEGRAL grant No. I/008/07.

\end{document}